\documentstyle[preprint,aps]{revtex}
\begin{document}
\draft
\title{Non-equilibrium dynamics in the random bond Ising chain:\\
A reminiscence of aging in spin glasses}
\author{H. Rieger, J. Kisker and M. Schreckenberg}
\address{
Institut f\"ur Theoretische Physik\\
Universit\"at zu K\"oln\\
50937 K\"oln,Germany
}
\date{\today}
\maketitle
\begin{abstract}
%*************************************
%
Results of extensive Monte-Carlo simulations that
investigate the out-of-equilibrium dynamics of the one-dimensional
Ising spin glass model with a Gaussian bond-distribution are presented.
At low enough temperatures a typical 
(interrupted) aging scenario is established as in two- and similar
to three-dimensional spin glass models. Since the underlying mechanism
is a slow domain-growth we study in detail spatial correlations
and the time-dependence of the domain- as well as kink-statistics.
We find that all correlation functions in time and in space as well as 
the domain-size probability distribution function obey simple
scaling laws.
%
%*************************************
\end{abstract}
\baselineskip22pt

\section{Introduction}

Non-equilibrium dynamics in real spin glasses, also known as
{\it aging}, has been a major focus of research interest of
experimentalists since many years \cite{age_review,CDW_age,super_age}. 
Quite recently, also theoreticians began to shift their attention from
the equilibrium statistical mechanics \cite{eq_review} to the 
out-of-equilibrium dynamics of various spin glass models
\cite{Bouch,JOA,Rieger,CuKu,PaMa,MeFr,Ritort,Thermo}.
It has been argued that aging is a characteristic feature of
the low temperature dynamics of spin glasses and it has already
been established in experiments \cite{CuMnfilm,Schins} 
and numerical simulations \cite{JOA,domain} investigating
two-dimensional spin glasses (which do not posses a finite
temperature phase transition) that a true spin glass phase is not a
necessary prerequisite for it to be observable on suitable
time-scales. The only differences remaining is then that aging is interrupted
at some maximal waiting time, which is equivalent to the finite,
but large, equilibration time at this temperature (see also \cite{CDW_age}
and \cite{Bouch}).

In this paper we will demonstrate that even in
a disordered but {\it non-frustrated} system a very similar scenario will
occur. The model we consider here is the one-dimensional Ising spin glass
with a Gaussian bond-distribution, which can be mapped, simply via a 
gauge-transformation of the spins,
onto a random, ferromagnetic Ising chain. Thus the ground state is
fully magnetized and the dynamics at low temperatures is 
characterized by (ferromagnetic) domain-growth, which is drastically 
slowed down by the presence of a huge number of metastable states
\cite{Chen,Colborne,Blundell}.  An important ingredience, however, is
a continuous, unbounded bond distribution, although some of the features
to be presented can be anticipated already in the one-dimensional 
$\pm J$ Ising spin glass model. The latter is isomorphic to the ferromagnetic
Ising chain, whose non-equilibrium dynamics is exactly solvable and
has been studied extensively \cite{Glauber,KoHi,Bray,Amar}. Only for
a continuous bond distribution one has a broad distribution of 
(free) energy barriers, which are responsible for the broad spectrum
of relaxation times.

The paper is organized as follows: The next section reports known
exact results for the autocorrelation function and the two-point
correlation function of $\pm J$ Ising chain. In section 3 we
present results of Monte-Carlo simulations of the random bond Ising
chain, which cover correlations in time and in space. A scaling
analysis along the lines sketched in section 2 is performed and
compared with the results of a recently proposed one-dimensional
domain-growth model \cite{Blundell}. Section 4 is 
devoted to a domain-size and -kink statistics that we obtained from our 
numerical simulations and here also a scaling analysis is performed.
Section 5 discusses the concept of an overlap length \cite{KoHi,FiHu} 
in the random bond Ising chain and summarizes our findings.

\section{The $\pm J$ chain}

In this section we recapitulate what is known about the
non-equilibrium dynamics of the $\pm J$ chain, which can be solved
exactly \cite{Glauber,KoHi,Bray,Amar,schreck}. On one hand we introduce
in this way the quantities that we intend to study for the random bond
Ising chain and on the other we shall see what kind scaling behavior can be
expected for these quantities.
The one-dimensional $\pm J$ Ising spin glass model (which is no spin
glass at all) is defined by the Hamiltonian 
\begin{equation}
H=-\sum_{i=1}^N J_i S_i S_{i+1}\;,
\label{hamil}
\end{equation}
where $S_i=\pm1$ are Ising spins and the bonds $J_i$ are chosen at random 
from a binary distribution
\begin{equation}
P(J_i)=\frac12[\delta(J_i-1)+\delta(J_i+1)]\;,
\label{binary}
\end{equation}
however, this quenched disorder can be removed (up to a negligible
boundary term) via the
Gauge-transformation $S_i'=S_i \prod_{k=1}^i {\rm sign}\,J_k$. 
Thus (\ref{hamil})
is equivalent to the simple ferromagnetic spin chain. All results
reported below are for $N\rightarrow\infty$. The dynamics 
is the usual Glauber-dynamics \cite{Glauber}, where each
spin is flipped with a probability
\begin{equation}
w(S_i\rightarrow-S_i)=\frac12[1-\tanh\,\beta(
J_{i-1}S_{i-1}+J_i S_{i+1})\,]\;.
\label{Glauber}
\end{equation}
The corresponding Master equation leads to a time dependent
probability distribution for spin configurations with which
expectation values are calculated (the initial distribution is
uniform, i.e.\ the at $t=0$ all spin are chosen at random).
Such a quantity for instance is the autocorrelation function
\cite{Glauber}
\begin{equation}
C(t,t_w)=\langle S_i(t_w) S_i(t+t_w)\rangle\;,
\label{corr}
\end{equation}
which will be studied numerically for the random bond chain in the
next section. At $T=0$ one obtains \cite{Glauber,Bray} 
\begin{equation}
C(t,t_w)=\frac2\pi {\rm arc\,\sin}
\left\{1+\frac12 \frac{t}{t_w}\right\}^{-1/2}\;.
\label{zero}
\end{equation}
Note that $C(t,t_w)$ is a function of $t/t_w$ only and such a scaling
behavior has been reported for the same quantity in various other 
spin glass models \cite{Bouch,Rieger,PaMa,Ritort,domain}. For illustration
and comparison with these models we have depicted in figure \ref{fig0}
a plot of $C(t,t_w)$ for various waiting times. Furthermore
one gets for short and for long times
\begin{equation}
C(t,t_w)\quad\approx\quad\left\{
\begin{array}{lcl}
1\,-\,\frac2\pi\,(t/t_w)^{1/2} & \quad{\rm for}\quad & t\ll t_w \;, \\
\frac{\sqrt{8}}{\pi}\,(t/t_w)^{-1/2} & \quad{\rm for}\quad & t\gg t_w \;.
\end{array}\right.
\label{corrasymp}
\end{equation}
For $T>0$ the time-dependence of $C(t,t_w)$ is determined by the
correlation length (which is infinite for $T=0$), which we consider now.
Spatial correlations are defined by 
\begin{equation}
G_T(r,t_w)=\langle S_i(t_w)S_{i+r}(t_w) \rangle\;.
\label{spatial}
\end{equation}
For $t_w\rightarrow\infty$ one has 
\begin{equation}
\lim_{t_w\rightarrow\infty}G_T(r,t_w)=G_T^{\rm eq}(r)=
\{\tanh(\beta J)\}^r
%
%\frac{\sum_{\underline{S}}\,S_i S_{i+r}e^{-\beta H(\underline{S})}}
%{\sum_{\underline{S}}\,e^{-\beta H(\underline{S})}}\;,
%
\label{spatialeq}
\end{equation}
which defines the equilibrium correlation length $\xi_{T,{\rm eq}}$ via
$\xi_{T,{\rm eq}}^{-1}=-\log\{\tanh(\beta J)\}$ \cite{Glauber}. 
It can be shown \cite{Bray} that for $1\ll t_w\ll\xi_{T,{\rm eq}}^2$
\begin{equation}
G_T(r,t_w)=\tilde{g}\left(\frac{r}{\xi_T(t_w)}\right)\;,
\label{spatialscale}
\end{equation}
where the characteristic length scale grows like
\begin{equation}
\xi_T(t_w)\propto\sqrt{t_w}
\label{sqrtgrowth}
\end{equation}
until it saturates for $t_w\rightarrow\xi_{T,{\rm eq}}^2$ and one recovers
$G_T(r,t_w)\rightarrow G_T^{\rm eq}(r)$.
Similarly, for $1\ll t_w\ll\xi_{T,{\rm eq}}$ one would expect instead of
(\ref{zero})
\begin{equation}
C_T(t,t_w)=\tilde{c}_T\left(\frac{t}{\tau(T,t_w)}\right)\;,
\label{corrscale}
\end{equation}
where the characteristic time scale grows like 
$\tau(T,t_w)\sim t_w$ for small temperatures,
but saturates at $\xi_{\rm eq}^2(T)$.
All the above results are very reminiscent of characteristic aging
phenomena observed in other spin glass models. The main difference
to these much more complex models lies in the waiting time dependence 
of the correlation length $\xi_T(t_w)$, which predicts a rather {\it fast}
domain growth, which makes it hard to observe them on macroscopic
time scales at non-zero temperatures. Furthermore, a significantly
different scaling behavior for e.g.\ $C(t,t_w)$ will be present in 
systems that are characterized by activated dynamics and 
logarithmic domain growth \cite{FiHu}.

\section{The random bond Ising chain}

The random bond Ising chain is defined in the same as the $\pm J$ Ising 
chain discussed in the last section up to a different bond-distribution:
Instead of (\ref{binary}) we consider here the Gaussian distribution 
of the random bonds

\begin{equation}
P(J_i)=\frac{1}{\sqrt{2\pi}}\,\exp\left(-\frac{J_i^2}{2}\right)\;,
\label{gauss}
\end{equation}
(see e.g.\ \cite{BhaYou} for the corresponding spin glass model in 
higher dimensions).
As before one can remove all minus-signs in the bonds by a 
gauge-transformation that leads to a purely ferromagnetic bond distribution 
$\tilde{P}(J_i)=2P(J_i)\theta(J_i)$. 

We performed Monte-Carlo simulations of the single-spin-flip Glauber
dynamics (\ref{Glauber}) with typical chain lengths of $N=10000$ 
(imposing periodic boundary conditions) and
averaged over 128 to 512 samples, i.e.\ realizations of the disorder.
Since we performed our simulations on a Parsytec GCel1024 parallel 
computer, using up to 256 different processors.
We constrained the system size to fit one system into the 
memory of one T805-transputer in order to avoid communication delays.
Alternatively we also could have made a few runs for chain-lengths
of $10^6$--$10^7$ and distribute chain-segments over several processors,
however, by varying the system size, we checked that our date were
not spoiled by finite-size effects.

The correlation functions (\ref{corr}) and (\ref{spatial})
have to be averaged over the quenched disorder. In addition,
in order to obtain better statistics we calculated instead of 
single-site/single-time quantities, appropriate space and time 
averages as
\begin{eqnarray}
C(t,t_w) & = &\frac1N \sum_{i=1}^N \frac{1}{\Delta t+1}
\sum_{t'=t}^{t+\Delta t}
[S_i(t'+t_w)S_i(t_w)]_{\rm av}\;,
\label{corr2}\\
G(r,t_w) & = & \frac1N \sum_{i=1}^N \frac{1}{\Delta t+1}
\sum_{t'=t}^{t+\Delta t}
[S_i(t'+t_w)S_{i+r}(t'+t_w)]_{\rm av}\;,
\label{spatial2}
\end{eqnarray}
with  suitable values for the time-window $\Delta t$
($1\le\Delta t\ll t$) and $[\cdots]_{\rm av}$
indicating the disorder average. If taken literally the expectation 
values in (\ref{corr}) and (\ref{spatial}) should be obtained via
different realizations of the thermal noise. However, we checked that
in using (\ref{corr2},\ref{spatial2}) we get the same results.

In figure \ref{fig1} we show our results for $C(t,t_w)$ at different temperatures.
The picture is equivalent to that obtained for the same quantity in the
two-dimensional Ising spin-glass \cite{domain} and, at low
temperatures, reminiscent of the three-dimensional
spin glass model \cite{Rieger} and the SK-model within the spin glass
phase \cite{Ritort}. This is what we would like to call interrupted 
aging as been observed experimentally in charge-density wave systems
\cite{CDW_age}. At higher temperatures the curves collapse for 
larger waiting times $t_w$, which means that the
system is equilibrated. The waiting time at which this happens is
the equilibration time $\tau_{\rm eq}$. At lower temperatures this
time scale is out of reach for the amount of CPU time that is
available to us, hence all curves seem to be shifted logarithmically
on the time axes by $t_w$.

Similar to what has been observed in other spin glass models
\cite{Bouch,Rieger,PaMa,Ritort,domain} we established the scaling
\begin{equation}
C(t,t_w)=
\tilde{c}_T\left(\frac{t}{\tau(T,t_w)}\right)\;,
\end{equation}
A scaling plot of this kind is depicted in figure \ref{fig2}, 
where the insert 
shows the characteristic time-scale $\tau(T,t_w)$. For small waiting
times at low temperatures it is, as expected, $\tau(T,t_w)\approx t_w$,
and it saturates for increasing $t_w$ at $\tau(T,t_w)=\tau_{\rm eq}(T)$.

Next we turn our attention to the spatial two-point correlation
function (\ref{spatial}). The (averaged) equilibrium correlation function can
easily be determined exactly as
\begin{equation}
G_{\rm eq}(r)=[\tanh(\beta J)]_{\rm av}^r
\label{eqg}
\end{equation}
(for a discussion about the differences between averaged and
typical or most probable values of the spatial correlations in this 
system see \cite{Derrida}) which yields an equilibrium correlation length
\begin{equation}
\xi_{\rm eq}^{-1}(T)=-\log\,[\tanh(\beta J)]_{\rm av}\propto T
\quad{\rm for\;T\ll1}\;.
\label{eqlength}
\end{equation}
The non-equilibrium spatial correlation function $G_T(r,t_w)$,
defined in (\ref{spatial}) respectively (\ref{spatial2}), will
approach (\ref{eqg}) at waiting times $t_w$ larger than the
equilibration time $\tau_{\rm eq}(T)$. This is shown in 
figure \ref{fig3} for two different temperatures, where at $T=0.4$ one 
observes indeed this data-collapse. The curves are approximate 
straight lines in a linear-log plot, which means that 
the spatial correlations decay exponentially with a characteristic
length scale $\tilde{\xi}_T(t_w)$. This length is roughly equal to
the length scale $\sigma_T(t_w)$ that yields a good data-collapse for a scaling
plot according to equation (\ref{spatialscale}), i.e.\
\begin{equation}
G(r,t_w)=\tilde{g}(r/\sigma(t_w))\;,
\end{equation}
which is shown in figure \ref{fig4}. The smoothest curve
for the waiting-time dependent length scale can be obtained via the
introduction of the (effective) correlation length
\begin{equation}
\xi_T(t_w)=\int_0^\infty dr\,G_T(r,t_w)\;,
\end{equation}
note, however, that for any $t_w$ and $T$ we have 
$\xi_T(t_w)\approx\sigma_T(t_w)\approx\tilde{\xi}_T(t_w)$.
The correlation length $\xi_T(t_w)$ is shown in the insert of 
figure \ref{fig4} within a log-log plot,
where one observes that this length scale grows much slower
than in the $\pm J$ chain, cf.\ (\ref{sqrtgrowth}). Similar to
what has been found in the two-dimensional spin glass model
\cite{domain} it seems that (see the curve for $T=0.2$)
$\xi_T(t_w)$ grows algebraically as
\begin{equation}
\xi_T(t_w)\propto t_w^{\alpha(T)}\qquad{\rm for}\quad t_w\ll\tau_{\rm eq}\;,
\label{corrgrowth}
\end{equation}
with a small, temperature-dependent exponent $\alpha(T)$ (note
that $\lim_{t_w\rightarrow\infty}\xi_T(t_w)=\xi_T^{\rm eq}(t_w)$,
so that again). 

The remanent magnetiziation $M_{\rm rem}(t)=C(t,0)$
decreases with increasing domain-size. If one assumes
$M_{\rm rem}(t)\propto\xi_T(t_w)^{-\lambda}$) as for instance 
in \cite{FiHu}, one obtains with (\ref{corrgrowth}) an algebraic
decay $M_{\rm rem}(t)\propto t^{-a(T)}$ as long as 
$\xi_T(t_w)\ll\xi_{rm eq}(T)$, which would concur with the prediction
\cite{Chen}. However, we find that the time dependence of the 
remanent magnetization can be better fitted to
a stretched exponential, similar to the lower bound
given in in \cite{Colborne}.

Starting from a one-dimensional domain growth model
the exact scaling function for the two-point correlations has been 
predicted in \cite{Blundell}. It is (for an unbounded bond-distribution)
$G(r,t_w)=\overline{c}(r/L_T(t_w))$ (similar to (\ref{spatialscale}))
with
\begin{equation}
\overline{c}(x)=\int_0^1 dy\,2y^2\exp(-2x/y)\log(y/(1-y))
\label{exform}
\end{equation}
and $L_T(t_w)$ the average domain-size after a waiting time $t_w$ at
temperature $T$. Note that $\overline{c}(x)\approx\exp(-2.19\,x)$
for $x\in[0,10]$ to great precision. By rescaling the variable
$x$ in (\ref{exform}) by a factor of roughly one half one gets a good
agreement with our data, as is shown in figure \ref{fig4}.
This means that the free parameter $L_0$ in \cite{Blundell}, which
is identified with the average domain-size, is approximately
twice as large as our correlation-length $L_0\approx 2\xi_T(t_w)$.
Note that due to the construction of the model \cite{Blundell}
no prediction about the time dependence of this length scale 
could be made.

\section{Domain-Statistics}

The slow domain growth in spin glasses in general \cite{FiHu,KoHi2} 
and in the random bond Ising chain in particular \cite{Chen,Colborne,KoHi}
is the reason for their glassy dynamics. This fact has already
been established numerically for instance in the site-diluted Ising 
model \cite{DIM}, the random field Ising model \cite{RFIM} and
the random bond ferromagnetic Ising model \cite{RBFIM} in two 
dimensions.
In contrast to the $\pm J$ chain, where
this slow dynamics is appropriately described by simple kink-diffusion
and -annihilation \cite{Amar} (which is also the ultimate reason for 
the "random-walk-result" (\ref{sqrtgrowth})), the quenched randomness
gives rise to a huge number of metastable states and (free) energy 
barriers of all sizes. Thus the mechanism of domain growth in the presence
of disorder is significantly modified and slowed down. However, it
is still describable by means of diffusion and annihilation of kinks
(broken bonds) in a random potential \cite{Chen}. Nevertheless
it has not been possible up to now to derive an analytically exact 
expression for the average domain-size as a function of the 
waiting time.

We calculated the time-dependent domain-size distribution during
our Monte-Carlo simulations in the following way (note that this
definition of domains is different from the cluster-definition given
in \cite{Coniglio}) : By recording 
the time averaged local magnetizations
\begin{equation}
m_i(t_w)=\frac{1}{t_w}\sum_{t=t_w}^{2t_w-1} S_i(t)
\end{equation}
for each sample we identified domains as connected segments 
of the chain in which all local magnetizations have the same sign.
The length of one domain is just the size of one segment.
Then we determine the number $n_l$ of domains of length $l$ in all 
samples ($\sigma=\sum_l n_l = N\cdot{\cal S}$, $N=$system size,
${\cal S}=$ number of samples) and get the probability 
$P(l,t_w)$ for a spin to be within a domain of length $l$ via
\begin{equation}
P_T(l,t_w)=l\,n_l(t_w)\,/\,\Sigma\;.
\end{equation}
In figure \ref{fig5} we depict the result for two different temperatures.
According to the scaling behavior found in the last section
the probability distribution for the domain-size should scale
in a similar way
\begin{equation}
P_T(l,t_w)\approx \frac{1}{L_T(t_w)}\,
\tilde{p}\,\left(\frac{l}{L_T(t_w)}\right)\;,
\end{equation}
An appropriate choice for $L_T(t_w)$ is 
the mean value or average domain-size $L(t_w)=\int dl\,l\,P_T(l,t_w)$,
a scaling plot is shown in figure \ref{fig6} and the insert shows $L_T(t_w)$
as a function of the waiting time. The shape of the curve is similar
to that of $\xi_T(t_w)$, although the average domain-size is much larger
than the correlation length. This is due to the definition of domains
that we have chosen: It is only the sign and not the magnitude 
of the local magnetizations that determines whether a spin belongs to a 
domain or not. If a spin has a rather small magnetization during a
time interval of length $t_w$ (which means that it has been 
dynamically "active" rather than frozen)
it contributes only marginally to the
correlation function, but fully to our domain-size. 
Therefore both quantities are expected to
be related to each other, but are not equal.

It is worth stressing that $P_T(l,t_w)$ is not Gaussian and that the
average $L_T(t_w)$ is dominated by the long tail of the domain-size
distribution. The typical domain-size, which is defined by the
length $l$, where $P_T(l,t_w)$ attains its maximum, is much smaller
than the average $L_T(t_w)$, however, it depends on the waiting time
in the same way as the average (different from what has been observed
in the random field Ising chain in a transverse field \cite{Fisher}).

The number of kinks $N_K(t_w)$ in the system, i.e.\ bonds between spins 
that are anti-parallel, is connected to the average domain length via
$L_T(t_w)=N/N_K(t_w)$. By inspecting the probability distribution
$K_T(J,t_w)$ for the kink-strength $J$ one obtains information about
the thermal energy of the chain. The ground state energy per spin of the
system is given by $E_0=-\int_0^\infty dJ\,P(J)$, for non-zero temperatures
(and for all waiting times $t_w$) broken bonds will increase the thermal 
energy by an amount $\Delta E_T(t_w)=\int_0^\infty dJ\,J\, K_T(J,t_w)$.
In figure \ref{fig7} we show the kink-strength distribution $K_T(J,t_w)$ for 
two temperatures. We note that, comparing it for $T=0.4$ with figure
\ref{fig5}, the kink-strength distribution seems to reach stationarity
(i.e.\ independence of $t_w$) much faster than the domain-size
distribution. Since the latter yields information about the magnetization
of the system, this is in full agreement with \cite{Chen}, where it
has already been noted, that the relaxation of the thermal energy is 
much faster than that of the magnetization.
The whole kink-strength distribution scales nicely with the 
characteristic energy scale $\delta E_T(t_w)$:
\begin{equation}
K_T(J,t_w)=
\frac{1}{\Delta E_T(t_w)}\tilde{k}\left(\frac{J}{\Delta E_T(t_w)}\right)\;,
\end{equation}
as is shown in figure \ref{fig8}.
The insert shows the waiting time dependence of the excess energy
$\Delta E_T(t_w)$ in a log-log plot, indicating an algebraic decay
over the range of time-scales considered (note that it has to
saturate at a non-vanishing value for non-zero temperatures,
which means that an algebraic decay cannot hold forever). This
is in agreement with what has been proposed in \cite{Chen}.

\section{Discussion}

Before summarizing our results we want to discuss the issue of
an overlap length \cite{KoHi,FiHu,KoHi2} in the random bond Ising chain.
It has been found that the spin-glass state is extremely 
sensitive to temperature- or field changes by an infinitesimal
amount \cite{BM}. This sensitivity is observable via the overlap
correlation function
\begin{equation}
\begin{array}{rcl}
K_{T,H,\Delta T,\Delta h}(r) & = &
[\langle S_i S_{i+r}\rangle_{T,h}^{\rm eq}
\langle S_i S_{i+r}\rangle_{T\pm\Delta T,H\pm\Delta h}^{\rm eq}]_{\rm av}\\
 & \propto &\exp\{-r/\lambda_{T,h}(\Delta T,\Delta h)\}\;,
\end{array}
\label{overlap}
\end{equation}
where the length scale $\lambda(\Delta T,\Delta h)$ is the so called
overlap length. For a system with a finite correlation-length at
temperature $T$ and field-strength $h$, e.g.\ two-dimensional
spin glasses or the random bond Ising chain, this is a quite
trivial observation. For instance in the $\pm J$ chain at $h=\Delta h=0$
(see section 2) it easy to see that
\begin{equation}
\lambda_T(\Delta T)=\{|\log\,\tanh\,[J/T]|+|\tanh\,[J/(T+\Delta T)]|\}^{-1}
\;.
\end{equation}
The nontrivial implication of (\ref{overlap}) is that it is 
expected to hold also {\it within} the spin glass phase of a true (e.g.\
higher-dimensional) spin glass, where the
correlation-length is infinite \cite{FiHu,KoHi2,Ritort_chaos}. The reason is a
full rearrangement of the minima of the free energy on length scales
larger than $\lambda$ by changing the external parameters like $T$ or $h$
only slightly. In \cite{FiHu} such a scenario has been introduced artificially
(however, motivated by renormalization arguments)
into the $\pm J$ chain via an explicit, stochastic temperature dependence
of the bonds $J_i$. 

In the one-dimensional spin glass model 
such a rearrangement of the minimum-energy
spin-configurations can be observed at zero temperature (where
the correlation length is indeed infinite) for a situation,
in which one switches on a small external field of strength $\Delta h$
\cite{Chen}.
In the presence of a field the ground state of the chain is fragmented 
into segments, in which all spins are aligned with one of the two possible 
configurations
with minimal energy at zero-field. In this process weak bonds have to
be broken and field energy can be gained and thus a new length scale 
$\lambda$ emerges. It can be shown \cite{Chen}
that the length of these segments, or the domain-size, depends on 
$\Delta h$ via
\begin{equation}
\lambda(\Delta h)\propto\Delta h^{2/3}\;,
\end{equation}
which can be identified with the overlap length (\ref{overlap})
for this particular situation $T=0$, $h=0$. At finite temperatures
the situation remains similar for small field changes \cite{Chen}.
Furthermore a generalization of the static overlap-correlation function
to a non-equilibrium situation is straightforward (see \cite{KoHi}),
but has not been investigated yet --- neither numerically nor analytically.

In conclusion we have presented a detailed numerical study of the
non-equilibrium dynamics of the random bond Ising chain. We have
shown that many of the aging phenomena reported in the literature
--- as for instance strong memory effects, $t/t_w$ scaling of the 
autocorrelation function and slow domain growth --- can also be
found in this non-frustrated model and also, although less pronounced,
in pure Ising models (see section 2, and also \cite{Janssen}, where
a similar scaling of the autocorelation function as in (\ref{corrscale})
has been derived within an $\varepsilon$-expansion around four dimensions
in a field theory for the non-equilibrium dynamics of ferromagnets).

We established simple
scaling of the two-point correlation function and showed that
the corresponding (temperature-independent) scaling function
concurs with that proposed very recently in a one-dimensional 
domain-growth model \cite{Blundell}. In addition we have
obtained the explicit time dependence of the characteristic
length scale. Also the domain-size distribution and the kink-strength
distribution exhibits simple scaling with an average domain size
and excess energy, respectively. The time dependence of the
latter quantities is algebraic as long as equilibration is not
achieved, which is very reminiscent of what has been observed in 
higher dimensional spin glass models \cite{Rieger,domain}.

Probably most people will agree that there is an essential
difference between the simple, albeit disordered, model considered
here and other spin glass models --- as for instance the
two-- and three-dimensional Edwards-Anderson model and the
Sherrington-Kirkpatrick model. Apart from the presence of frustration
the most significant feature
of the latter, as compared to the random bond Ising chain, is
the occurrence of replica symmetry breaking. It seems that this
equilibrium feature of the SK-model has also interesting consequences
for the non-equilibrium dynamics \cite{CuKu,MeFr}. Nothing of this
will ever be observed in the random bond Ising chain. However,
many experimental observations on real spin glasses are
reminiscent of what we have reported here. Thus, as has already
been noted in \cite{KoHi}, it seems to us to be very useful to
look closer into one-dimensional models to catch some of the
physics in three dimensions.

\section[*]{Acknowledgement}

We would like to thank the Center of Parallel Computing (ZPR) in K\"oln 
for the generous allocation of computing time on the transputer cluster
Parsytec--GCel1024. This work was performed within the SFB 341
K\"oln--Aachen--J\"ulich.

\begin{figure}
\caption{The autocorrelation function $C(t,t_w)$ for the $\pm J$ Ising
chain at zero temperature, obtained from the exact result (5).}
\label{fig0}
\end{figure}

\begin{figure}
\caption{Autocorrelation function $C(t,t_w)$ at different temperatures.}
\label{fig1}
\end{figure}

\begin{figure}
\caption{Scaling plot of $C(t,t_w)$ versus $t/\tau_T(t_w)$ at 
temperature $T=0.1$.
the insert shows the characteristic time-scale $\tau_T(t_w)$ in a 
log-log plot} \label{fig2}
\end{figure}

\begin{figure}
\caption{Two-point correlation function $G_T(r,t_w)$ for different 
temperatures ($T=0.2$ left and $T=0.4$ right)}
\label{fig3}
\end{figure}

\begin{figure}
\caption{Scaling-plot of $G_T(r,t_w)$ versus $r/\xi_T(t_w)$ for $T=0.4$.
The full line is the theoretical prediction scaled with a factor $0.46$
(see text). The corresponding scaling plot for $T=0.4$ yields an
identical scaling function.
The insert shows the correlation length $\xi_T(t_w)$ in a log-log plot.}
\label{fig4}
\end{figure}

\begin{figure}
\caption{Domain-size distribution function $P_T(l,t_w)$ for 
different temperatures($T=0.2$ left and $T=0.4$ right).}
\label{fig5}
\end{figure}

\begin{figure}
\caption{Scaling-plot of $P_T(l,t_w)$ versus $L_T(t_w)$ at $T=0.2$. 
The insert shows the average domain-size $L_T(t_w)$ in a log-log plot.
The corresponding scaling plot for $T=0.4$ yields an
identical scaling function.}
\label{fig6}
\end{figure}

\begin{figure}
\caption{Kink-strength distribution $K_T(J,t_w)$ for different 
temperatures ($T=0.2$ left and $T=0.4$ right).}
\label{fig7}
\end{figure}

\begin{figure}
\caption{Scaling-plot of $K_T(j,t_w)$ versus $\Delta E_T(t_w)$ for $T=0.2$.
The insert shows $\Delta E_T(t_w)$ in dependence of $t_w$ in a log-log plot.
The corresponding scaling plot for $T=0.4$ yields an
identical scaling function.}
\label{fig8}
\end{figure}

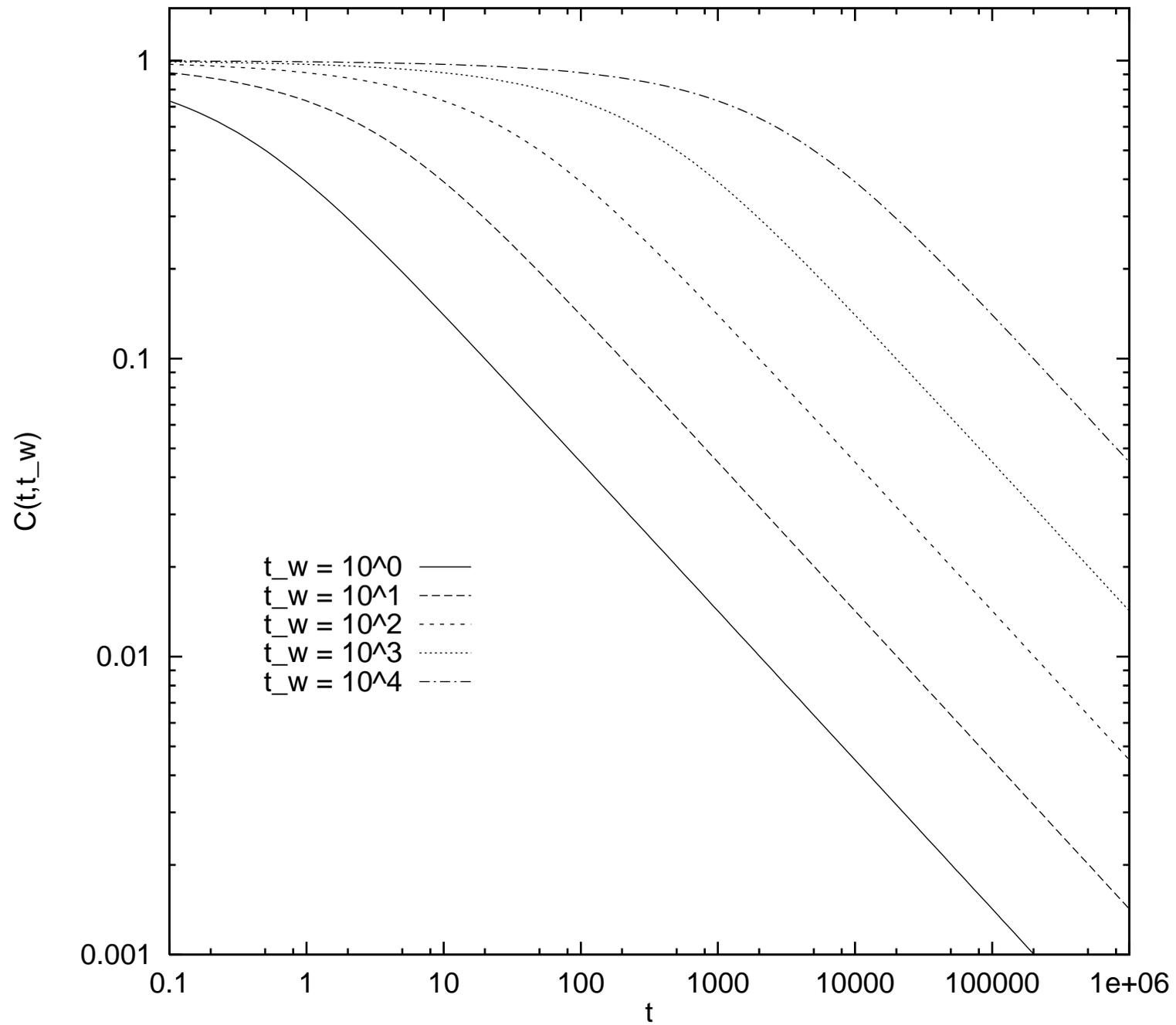
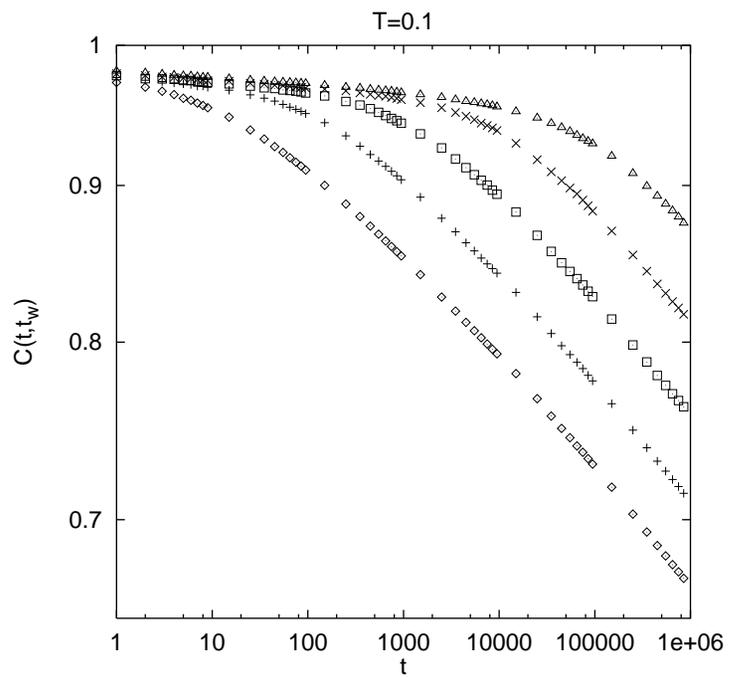
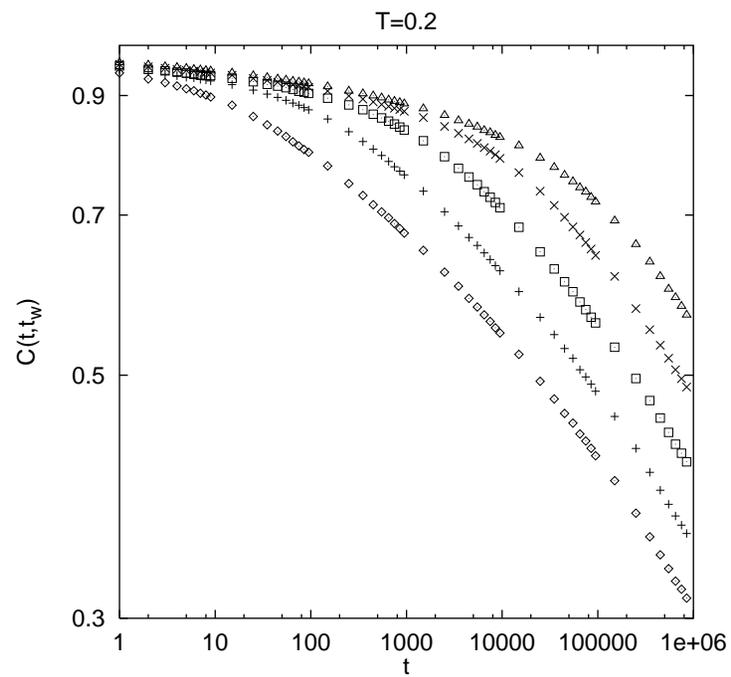
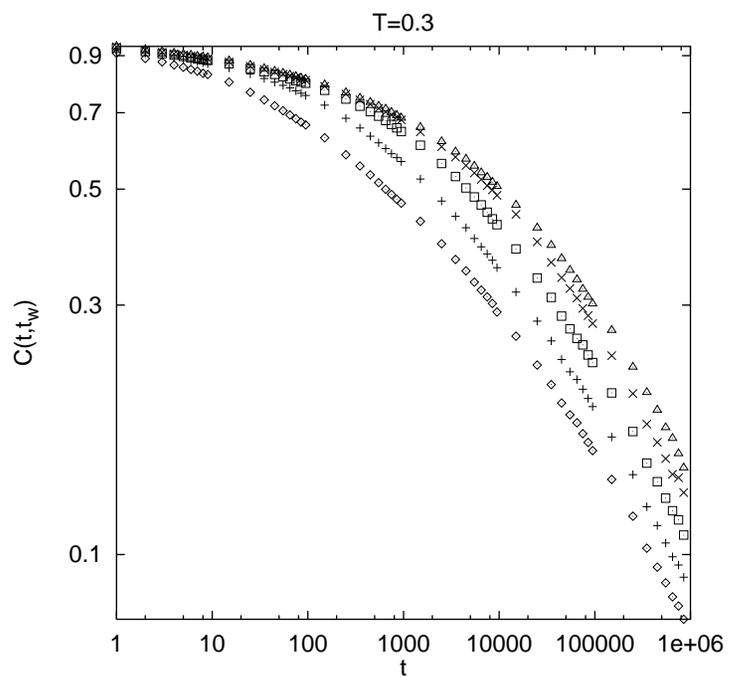
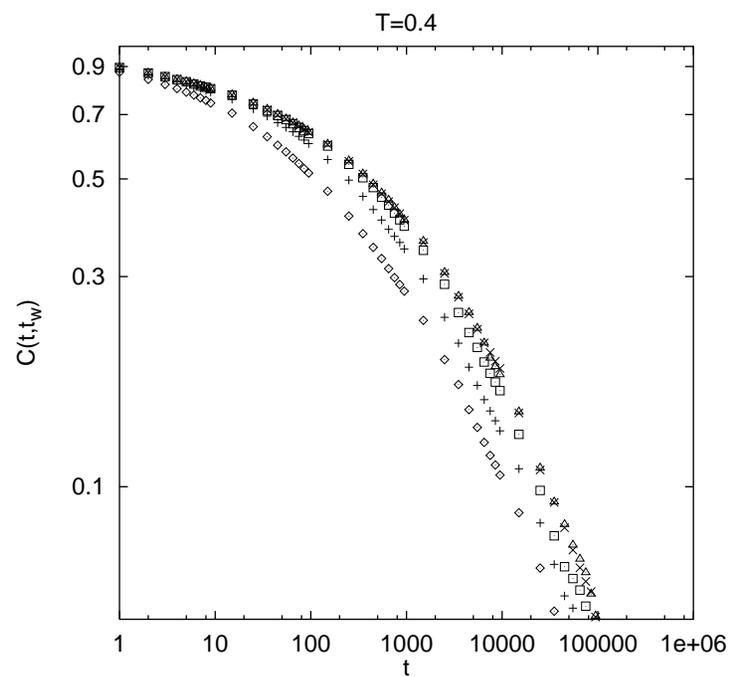
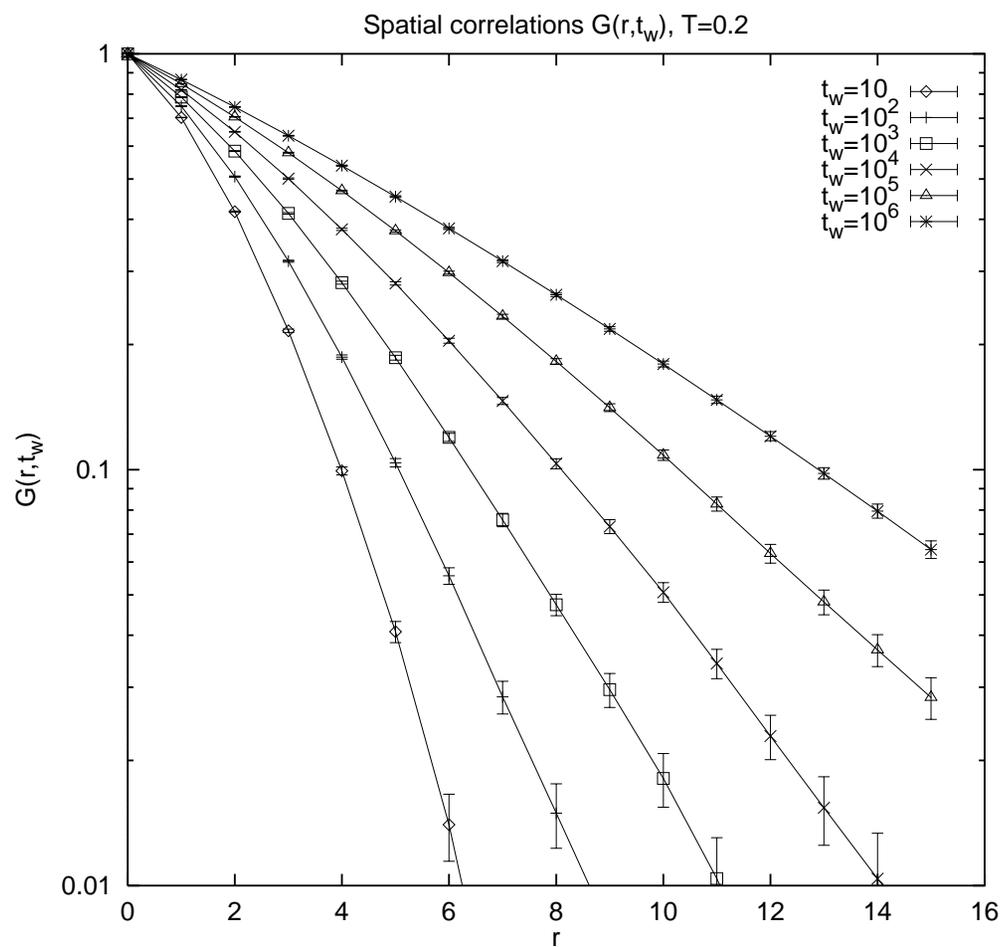
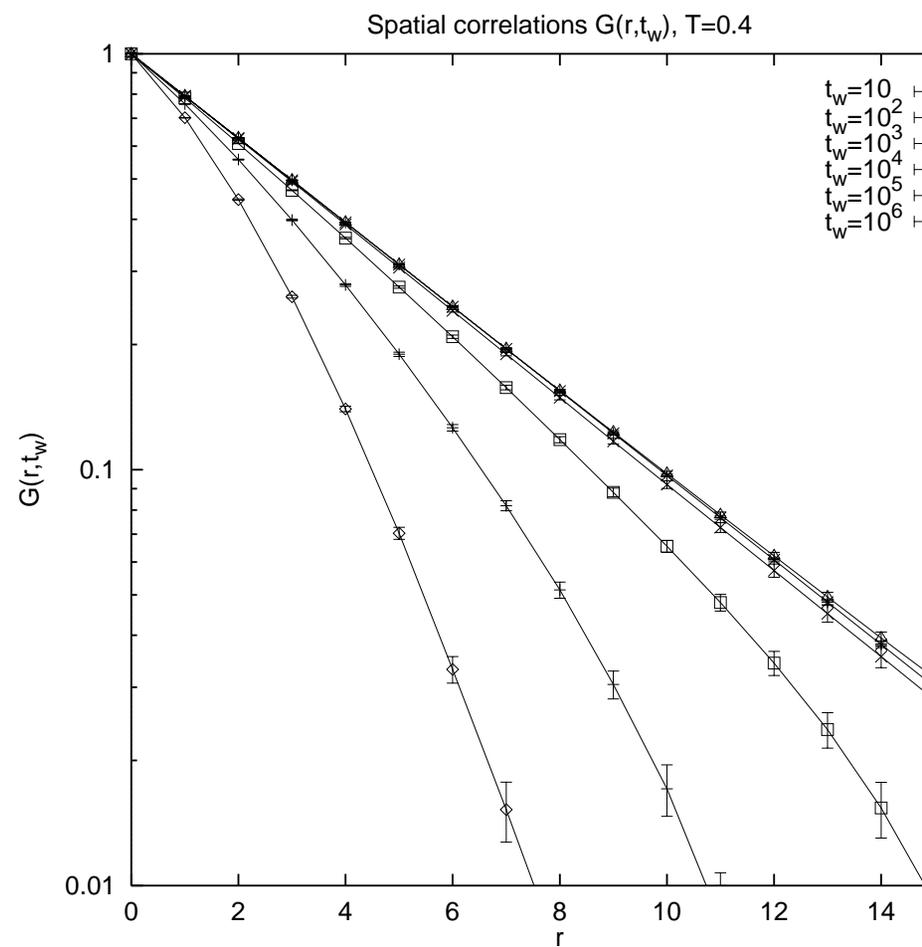
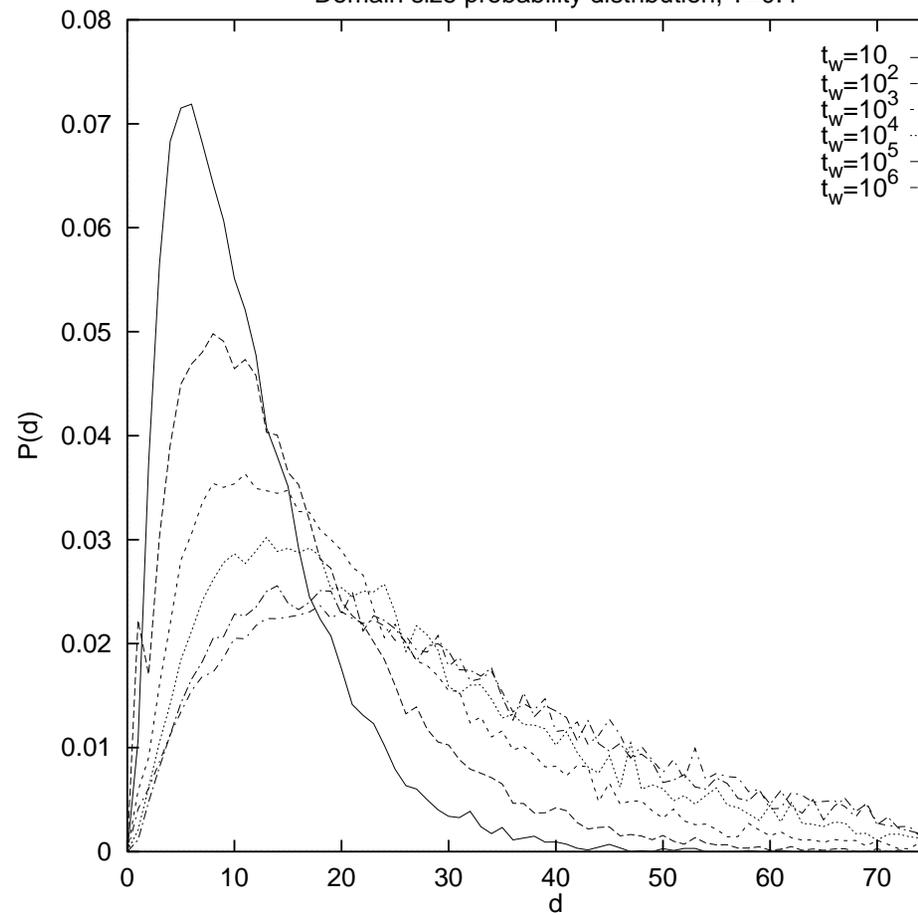
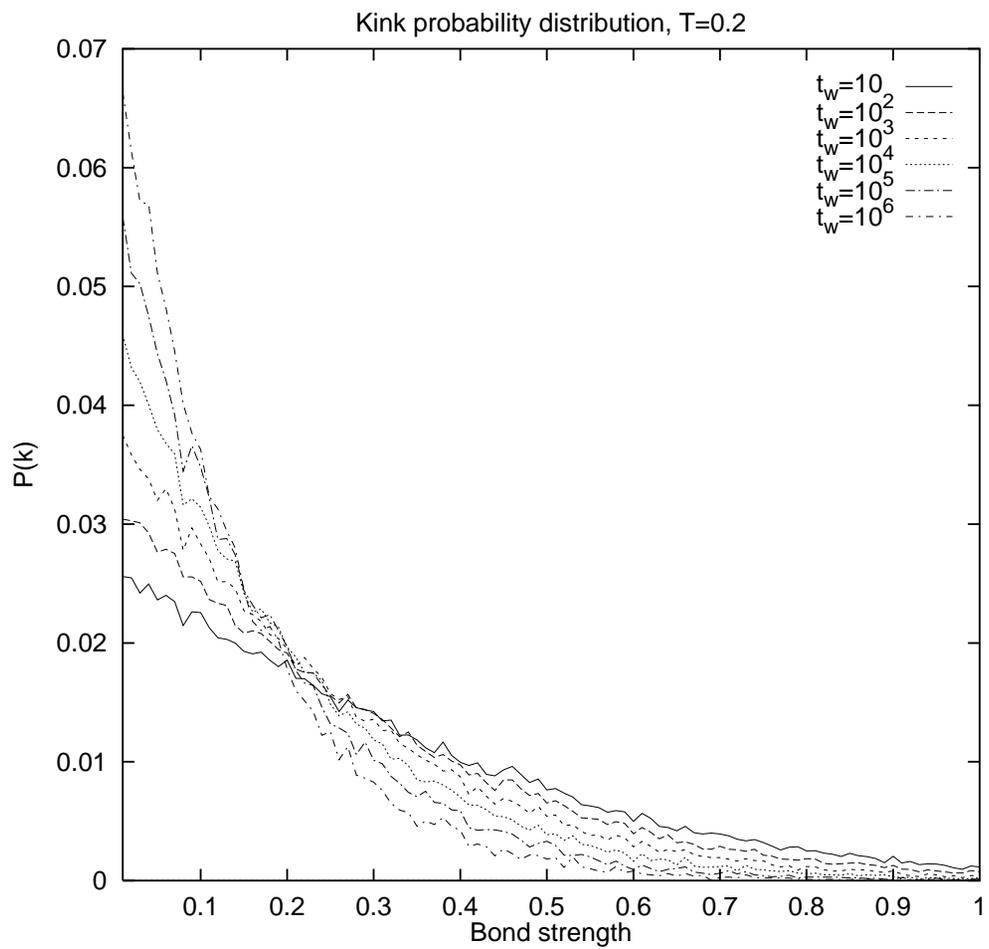
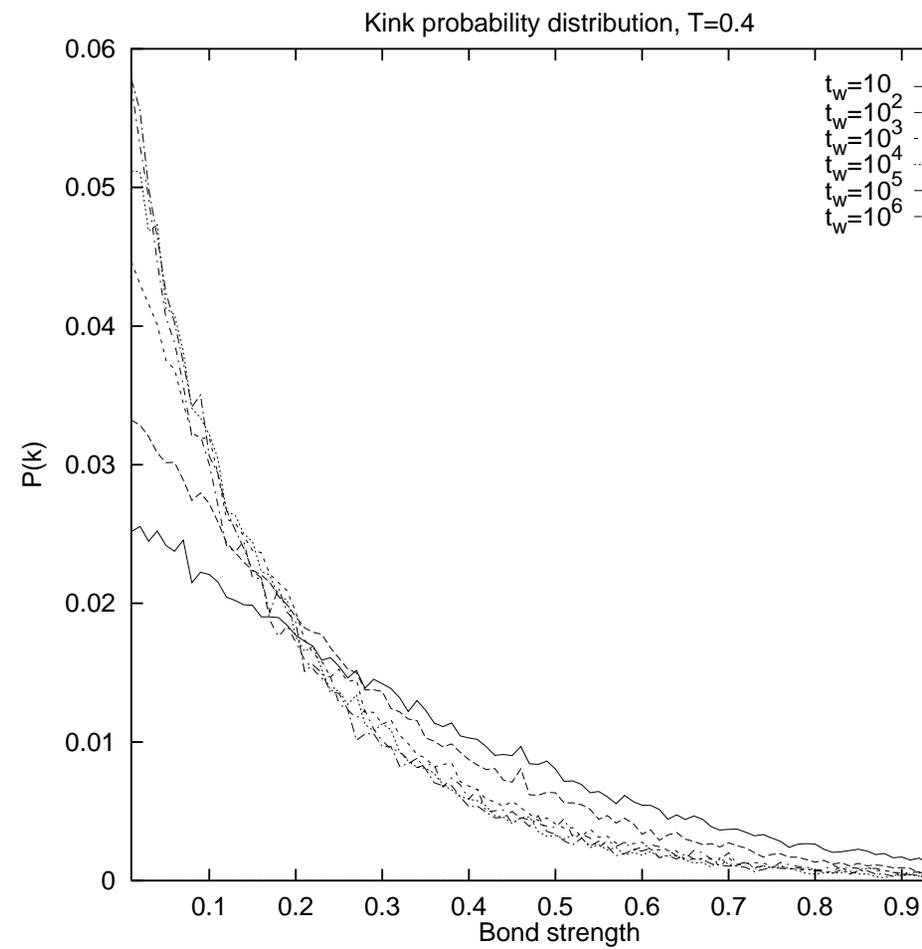
\end{document}